\documentclass[
aps,
pra,
article
longbibliography,
amssymb,
showpacs,
superscriptaddress,
notitlepage,
onecolumn
]{revtex4-2}
\usepackage{bm}
\usepackage{graphicx}
\usepackage{hyperref}
\usepackage{dcolumn}
\usepackage{braket}
\usepackage{amsmath}
\usepackage{color}
\usepackage{mathtools,amssymb}
\usepackage{soul}
\usepackage{float}
\usepackage{amssymb}
\usepackage{esint}
\usepackage{mathtools}
\usepackage{physics}
\usepackage{makecell}
\usepackage{multirow}
\usepackage{array}
\usepackage{geometry}
\usepackage{url}
\usepackage{adjustbox}
\usepackage{minted}
\usepackage{rotating}
\usepackage{listings}
\usepackage{threeparttable,booktabs}
\usepackage{fontawesome}
\usepackage{comment}
\usepackage{cleveref}
\usepackage{cuted}
\usepackage{tikz}
\usepackage{titletoc} 

\usepackage{svg}
\usepackage{url}
\usepackage{textcomp}
\newcommand{\micro}{\textmu}


\newcommand{\SMsec}[1]{Appendix~\ref{#1}}

\begin{document}
\title{Broadband spatiospectral mode conversion via four-wave mixing}
\author{Helaman Flores}\email{floresh2@mit.edu}
\affiliation{Research Laboratory of Electronics, Massachusetts Institute of Technology, Cambridge, Massachusetts 02139, USA}
\author{Mahmoud Jalali Mehrabad}
\affiliation{Research Laboratory of Electronics, Massachusetts Institute of Technology, Cambridge, Massachusetts 02139, USA}
\author{Siavash Mirzaei-Ghormish}
\affiliation{Department of Electrical and Computer Engineering, Brigham Young University, Provo, Utah 84602, USA}
\author{Ryan M. Camacho}
\affiliation{Department of Electrical and Computer Engineering, Brigham Young University, Provo, Utah 84602, USA}
\author{Dirk Englund}\email{englund@mit.edu}
\affiliation{Research Laboratory of Electronics, Massachusetts Institute of Technology, Cambridge, Massachusetts 02139, USA}
%
\begin{abstract} 
We introduce a framework for scalable and broadband free-space phase-matched four-wave mixing in ring resonators. This method for four-wave mixing reduces the complexity of coupling an emitter to a quantum network by combining the spatial and spectral interfaces between them into one nonlinear optical process. The device is compliant with current heterogeneous integration capabilities and supports frequency phase matching across approximately 165 nm. We outline a fabrication-ready diamond-on-insulator pathway towards modular unit cells that natively bridge visible color centers to the infrared spectrum for scalable quantum networks. We also present and analyze an end-to-end framework for considering single-photon coupling efficiency from a color center to a quantum network. This framework represents a step forwards in analyzing and reducing system-scale losses in a spin-photon interface.
\end{abstract}
\maketitle
%
\section{Introduction}\label{m_Intro}
In order to share computing power between distant quantum computers, a quantum network, complete with quantum repeaters, is necessary~\cite{Azuma:2023,lu2019chip,wang2025large,li2025down}. Many promising platforms for a quantum repeater have photonic interfaces in the visible spectrum, including trapped ions, neutral atoms, and color centers in diamond~\cite{Ruf:2021,Covey2023,Ritter2012,Hoffman:2012,Stephenson:2020,Krutyanskiy:2023}. However, in order to achieve long-distance transport of an entangled photon, light from any such platform requires Quantum Frequency Conversion (QFC) into the telecom band, where optical fibers operate with lower loss~\cite{Ruf:2021, Bock2018, Laneve2025, Strobel2025, singh2019quantum, huang2021hybrid, raghunathan2025telecom, li2016efficient}. Thus, the conversion between visible and infrared wavelengths is an integral problem in the field of quantum networking.

A node in a future quantum network will also require integration of thousands of qubits on one chip. Given the recent developments of high-speed spatial light modulators (SLM) as a switch, multiplexing free-space coupled qubits shows increasing promise to meet the needs of large-scale qubit integration~\cite{duan2021vertically,Peng:19, Panuski2022, Ding:2023}.

Most implementations of quantum frequency conversion involve either $\chi^{(2)}$ sum- and difference-frequency generation or $\chi^{(3)}$ Bragg-scattering four-wave mixing (BS-FWM). BS-FWM has enabled efficient, low-noise spectral translation in fibers, waveguides, and microresonators, including operation at the single-photon level and across visible--telecom wavelength separations~\cite{McKinstrie:05,li2016efficient,singh2019quantum,heuck2019unidirectional,raghunathan2025telecom}. In representative integrated BS-FWM architectures, however, the signal and idler are both guided or cavity-confined modes, and the converted idler subsequently couples to an access waveguide~\cite{li2016efficient,heuck2019unidirectional,raghunathan2025telecom}. Frequency translation and spatial out-coupling are therefore engineered as separate functions.

We propose a scheme by which a two-pump $\chi^{(3)}$ interaction can convert a cavity-bound visible photon directly into a deliberately radiative infrared mode, so that spectral conversion and free-space extraction occur in the same nonlinear process. The conversion requires neither structured pump illumination nor an SLM; the output is a fixed, on-axis channel, while the second pump supplies spectral freedom through the choice of resonant pump-mode pairs. One concurrent manuscript explores a different design axis utilizing the $\chi^{(2)}$ nonlinearity: a single structured pump writes a reconfigurable virtual Bragg grating that programs the spatial and orbital-angular-momentum content of the emitted field in situ~\cite{mirzaei2025programmable}. In contrast, our investigation simplifies a spin-photon interface to minimize the amount of fabrication steps and photonic interfaces possible. We then quantify and minimize each potential system level loss by simulating optimal operating points for an experiment.

Diamond is particularly suited to the needs of a quantum repeater because the color-center emitter and nonlinear resonator can occupy the same device, reducing the need for an interface to a separate chip for QFC~\cite{mccutcheon2009broadband}. Diamond also provides an appreciable $\chi^{(3)}$ nonlinearity, with nonlinear refractive index $n_2=8.2\cdot10^{-20}$~$\mathrm{m^2W^{-1}}$. Other prior works have demonstrated high-quality-factor diamond resonators~\cite{hausmann2014diamond,Ding2024}, while diamond-chiplet transfer techniques provide a pathway to heterogeneous integration~\cite{Starling2023,Pholsen:25}.

In this work, we introduce a fabrication-oriented diamond ring resonator that uses free-space phase-matched four-wave mixing (FFWM) as an end-to-end spin--photon interface architecture. Our contributions are threefold. First, we design the in-plane angular momenta of the signal and two pumps so that the induced $P^{(3)}$ drives an $m_{idl}=0$, low-$Q$ idler mode that radiates normal to the ring; the nonlinear interaction itself therefore acts as the output coupler. Second, a fixed ring geometry supports multiple discrete resonant pump combinations whose out-of-plane phase-matching window spans approximately 165~nm and is centered near the telecom O-band, without geometric reconfiguration or thermal tuning of the resonator. Third, we develop an end-to-end emitter-to-fiber model that includes nonradiative and zero-phonon-line branching, Purcell-enhanced loading of the signal mode, pump-dependent signal--idler coupling, the engineered idler radiation rate, and free-space collection. This system-level treatment goes beyond conversion between externally injected guided modes and identifies the operating point that maximizes the probability of retrieving a telecom photon following a single color-center decay.

\begin{figure*}
    \centering
\includegraphics[width=0.9\linewidth]{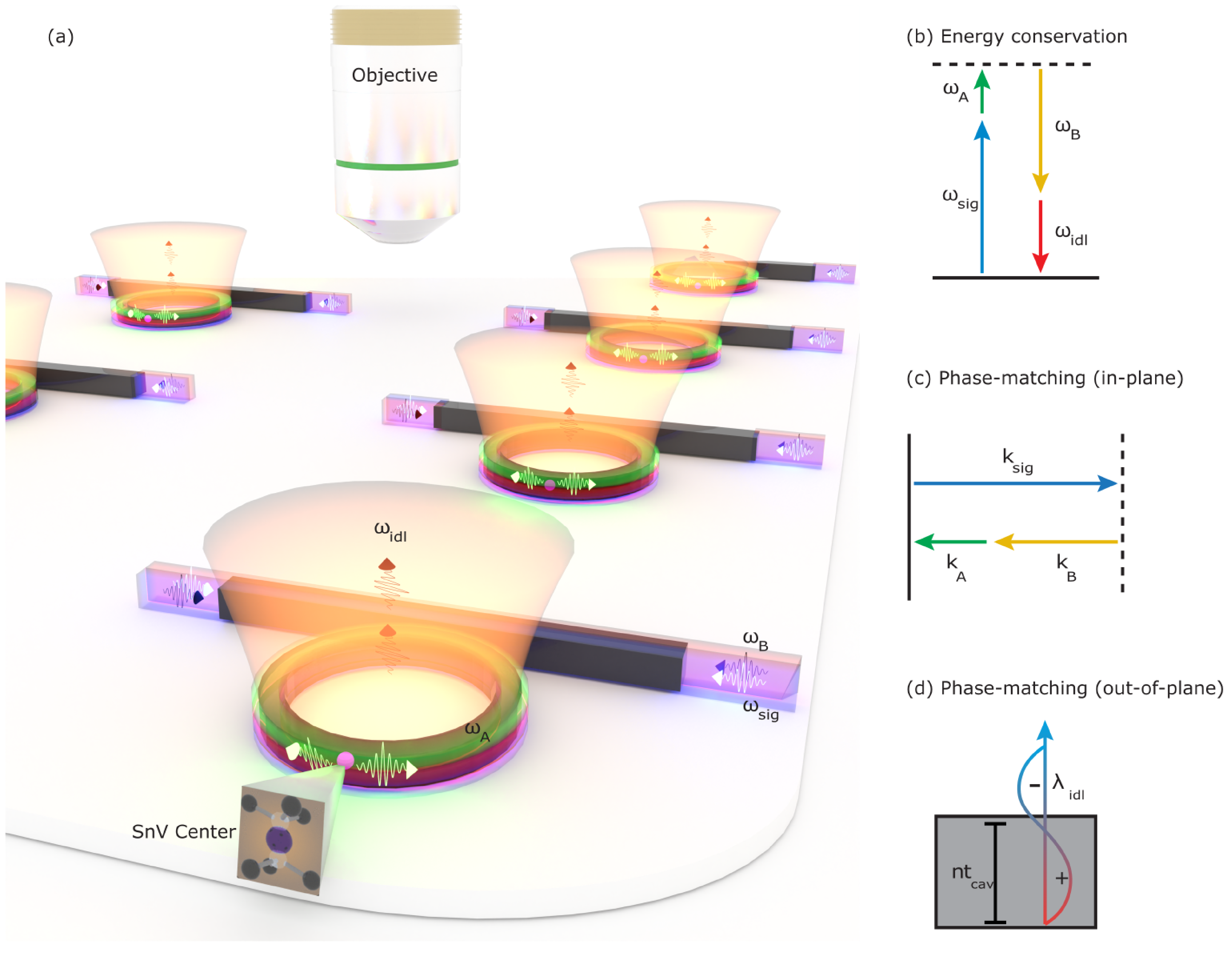}
\caption{ (a) Light couples into a diamond resonator from a silicon nitride bus waveguide. The bus sits next to the ring, with a back reflector embedded in the oxide. The signal emits into the ring from a color center embedded in the diamond. Light converts to the infrared using the $\chi^3$ nonlinearity in diamond, and emits into free space for collection into an objective lens that connects to an optical fiber. This forms a unit cell for many copies of the same device that can be addressed by an objective lens. (b) The energy conservation condition for QFC in Bragg scattering mode. (c) The phase matching condition in the plane of the cavity. (d) The phase matching condition in the free space direction, out of the plane of the cavity.}
    \label{fig:Fig1}
\end{figure*}

The ring resonator couples to a single silicon nitride bus waveguide, depicted in Figure~\ref{fig:Fig1}a, and emits an infrared idler photon (1301 nm) into free space, converted from a visible signal photon (615 nm) from an emitter embedded in the diamond. Pump photons enter from the bus waveguide. A back reflector breaks the symmetry of the emission scheme, sending output idler photons in the $+z$ direction, which are then collected by an objective lens before passing through a S-waveplate and coupling to single-mode fiber. The device is defined by its ring radius $r$, width $w$, thickness $t_{cav}$, as well as the gap $d_{ref}$ between the bottom of the diamond and the top of the back reflector. The idler photon generated out of the cavity plane traverses the optical path length $\Lambda=n(\omega_{idl})t_{cav}$, where \(n(\omega_{idler})\) is the refractive index of the diamond. The polarization $P^{(3)}=\chi^{(3)}E_{sig}E_{A}E_{B}^*$ has a uniform phase along $\Lambda$. Thus, the phase accumulation of the idler photon as it traverses $\Lambda$ must be close to or less than $\pi$, as depicted in Figure~\ref{fig:Fig1}d, to enable constructive interference between the idler and the $P^{(3)}$.

We define the efficiency of our system as the probability of success for retrieving a single photon from the color center into a quantum network upon the decay of the emitter from the excited state, 
\begin{equation}
\begin{aligned} 
\eta=\eta_{spatial}\eta_{idler},
\end{aligned}
\end{equation}
where $\eta_{spatial}$ is the overlap of the electric field of the idler photon in the far-field $E_{ff}$ with a Gaussian beam, and $\eta_{idler}$ is the probability that a photon at the Zero-Phonon Line (ZPL) is retrieved from the emitter's decay and converted to the idler mode.
%
\section{Phase Matching Condition}\label{m_PM}
By inducing $P^{(3)}$ in our nonlinear medium that overlaps well with the paraxial wave equation solutions, we can shape propagation into the far-field at our target frequency $\omega_{idl}$. This stems from the fact that far-field propagation of an unconfined near field mode obeys the paraxial wave equation, and a mode without any momentum in the plane of a ring resonator will be nearly unconfined~\cite{Haus_2004}. We define the FPM conditions by sorting the involved waves into the sets $A_{p}$ and $A_{o}$, such that $A_{p} = \{sig, A, B\}$, $A_o={idl}$, representing whether the mode is confined in the cavity plane or traveling out of plane, respectively. Next, we constrain the phase-matching condition in Figure \ref{fig:Fig1}c to require zero in-plane idler momentum,
\begin{equation}
\begin{aligned}
\Delta k_{p}  = \sum_{i\in A_{p}}c_{i}n_{eff,i}p_{i}k_{0,i} = 0,
\end{aligned}
\end{equation}
where $k_{0,i}$ is the wave vector, $n_{eff,i}$ is the effective index for the wave in the cavity, and $p_{i}\in{-1,1}$ indicates the direction of propagation. When $\Delta k_{p}=0$, the idler photon constructively interferes in the far-field orthogonal to the plane of the cavity. In a ring resonator, $\Delta k_{p}=0$ is analogous to exciting the $m=0$ azimuthal order mode with $P^{(3)}$. Others have used this mode before as a static electric field~\cite{Lu2021}. However, we study its properties at non-zero frequencies.

We consider the energy conservation condition displayed in Figure \ref{fig:Fig1}b,
\begin{equation}
\begin{aligned}
\sum_{i\in(A_p\cup A_o)}c_i\omega_i=0,
\end{aligned}
\end{equation}
where the parameter $c\in{-1,1}$ determines the sign of each mode in the FWM interaction. We select $c_{sig}=1, c_{A}=1, c_B=-1, c_{idl}=-1$. This Bragg scattering scheme enables transfer of power from the signal to the idler without vacuum field amplification~\cite{McKinstrie:05, heuck2019unidirectional}. An emitter such as the SnV center is placed in the diamond resonator, coupling to a whispering gallery mode (WGM) with azimuthal number $m_{sig}$ and symmetric emission given by $p_{sig}=\pm1$. The pumps must excite both the clockwise and counter-clockwise direction $p_{A}=p_B=\pm1$ of the WGM, with azimuthal mode numbers $m_A$ and $m_B$ to match the emitter symmetry. The resulting FPM equations are
\begin{equation}
\begin{aligned} 
0=\omega_{sig}+\omega_A-\omega_B-\omega_{idl},
\end{aligned}
\end{equation}
\begin{equation}
\begin{aligned}
0=m_{sig}-m_A-m_B.
\label{pm_cond}
\end{aligned}
\end{equation}
Additional insight into competing nonlinear interactions and noise processes is given in~\SMsec{sm:noise}. 

\begin{figure*}
    \centering
    \includegraphics[width=0.8\linewidth]{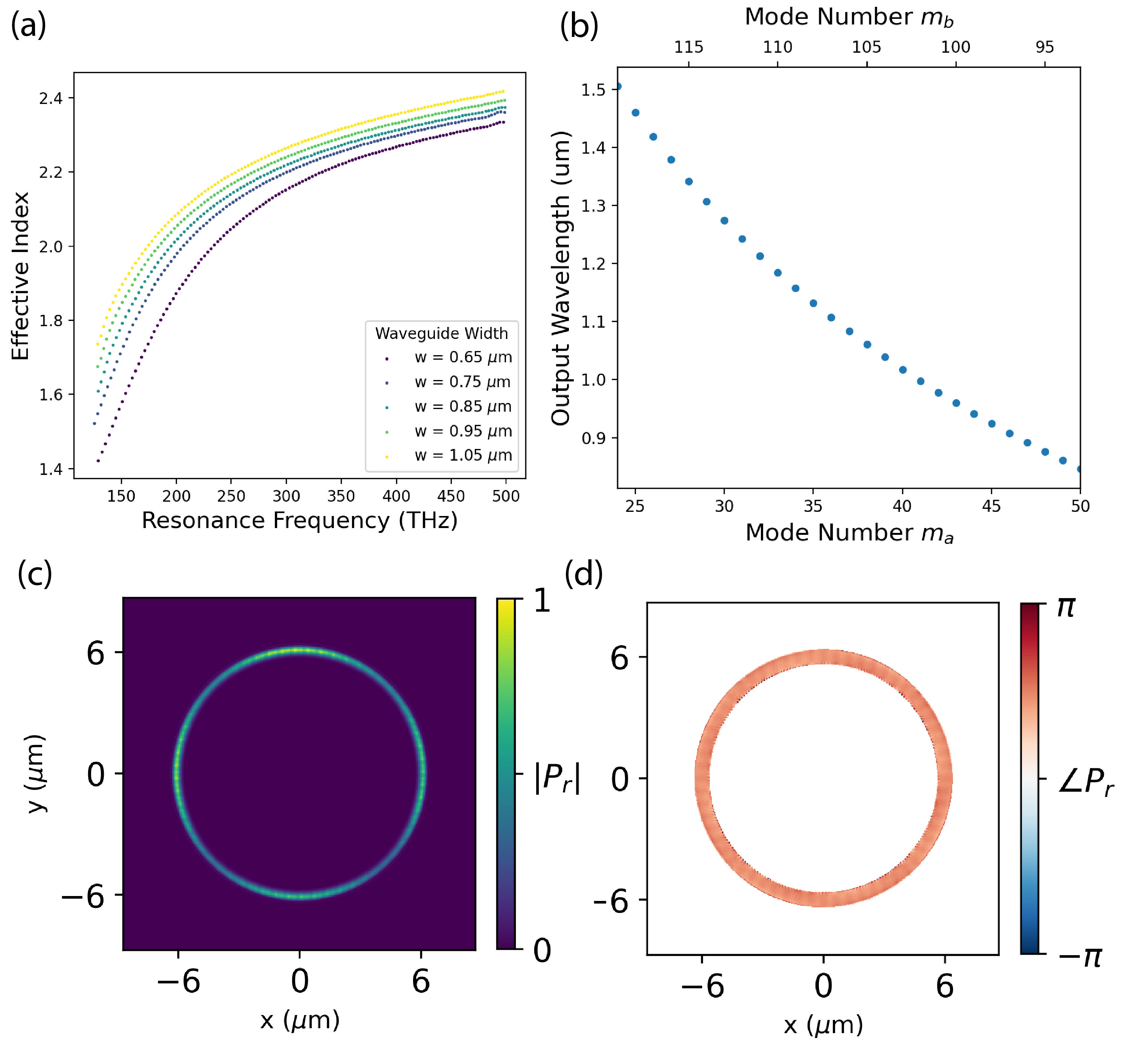}
    \caption{(a) The effective refractive index for discretized mode numbers of the ring resonator from mode solver data, plotted for multiple ring widths. (b) Output wavelengths that satisfy the in-plane phase matching condition in Equation~\ref{pm_cond}. (c) Polarization in the $\hat{r}$ direction obtained from FDTD simulation. (d) The phase of the output polarization in the $\hat{r}$ direction, confirming that the FPM condition $m_{idl}=0$ is fulfilled.}
    \label{fig:Fig2}
\end{figure*}

We solve for multiple idler wavelengths that satisfy these equations using a combination of FDTD and 2D mode solver simulations, reporting the results in Figure \ref{fig:Fig2}\cite{flores_tidy3d2026}. By fitting the $n_{eff}$ of a ring mode across a broad range of wavelengths we can then find the discretized mode numbers $m_i=2\pi rn_{eff}/\lambda_0$, plotted in Figure \ref{fig:Fig2}a. The idler wavelength can be calculated from the set of pump modes that fulfill Equation~\ref {pm_cond}, as shown in Figure~\ref{fig:Fig2}b. Each point in this figure represents an in-plane phase-matched solution for QFC of a photon from 615 nm to the given idler wavelength, without the need for any resonance tuning. However, the out of plane phase mismatch will increase at shorter wavelengths, as the photon will near a $2\pi$ phase shift relative to the nonlinear polarization on its journey to exit the cavity. This phase mismatch increases the decay rate of the idler mode, which in turn increases the pump power required to convert a photon from the signal to the idler mode.

Inside of a single ring, we find multiple solutions for Equation~\ref{pm_cond} corresponding to distinct output idler wavelengths that span over 500 nm in the infrared regime. However, the quality factor of the mode excited by $P^{(3)}$ indicates that the out of plane phase matching condition will be fulfilled for a limited subset spanning approximately 165 nm in the O-band. When conditions for Equation~\ref{pm_cond} are met, an in plane cross-section of the $P^{(3)}$ inside of the ring, shown in Figure~\ref{fig:Fig2}c-d, has high overlap with a radially polarized beam with azimuthal mode number $m=0$. This is a solution to the paraxial wave equation that remains confined to a low numerical aperture~\cite{Youngworth:00}.

%
\section{Simulation and Spatial Efficiency}\label{m_Design}
\begin{figure*}
    \centering
    \includegraphics[width=0.99\linewidth]{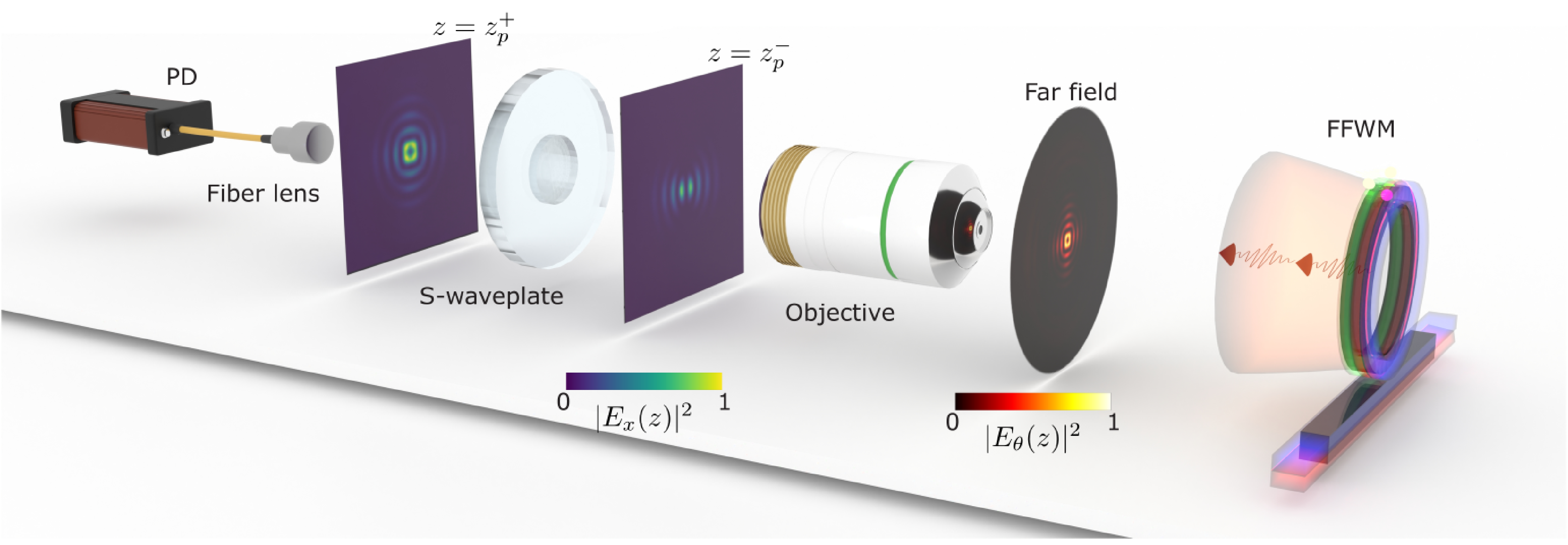}
    \caption{The ring emits the idler into the far-field through FFWM. We calculate the transverse idler electric field profile after the objective using the Debye-Wolf integral at $z_p^-$, just before the S-waveplate. The idler field after S-waveplate transformation at $z_p^+$ can be collected with efficiency $\eta_{spatial}$ into the fiber lens. See~\SMsec{sm:Spatial} for detailed field profile simulations. }
    \label{fig:Fig3}
\end{figure*}

We obtain the 3D profile of $P^{(3)}$ from the electric field of 3 FDTD simulations for the signal and pump modes, described in further detail in ~\SMsec{sm:RingProfiles}. For simplicity, we exclude the bus waveguide from the simulation of the signal and pumps, injecting directly into the WGM. We define the idler mode of the FWM interaction as equivalent to $P^{(3)}$, which has a similar mode volume to the signal and pump fields. To determine the far-field and quality factor of the idler mode, we inject $P^{(3)}$ into a final FDTD simulation as a perturbative current source, measuring the fields inside the cavity as well as its emission. The far-field of this mode, shown in Figure~\ref {fig:Fig3}, can be focused by an objective lens. We calculate the electric field at some distance $z_p^-$ from the focal plane of this objective lens using the Debye-Wolf integral, with a numerical aperture of 0.82, which captures 72\% of the power in the far-field~\cite{Sherif:08}. We sweep of $z_p$, apply the Jones matrix for an S-waveplate to the electric field to retrieve the field at $z_p^+$, then fit the output field to a Gaussian, resulting in a maximum spatial power overlap of 24\% at $z_p=27.2$  \micro m~\cite{BhargavaRam:17}. Assuming 90\% transmission through the coated objective and S-waveplate, we find $\eta_{spatial}=0.21$.

Determining the amount of constructive interference between $P^{(3)}$ and the idler photon enables us to calculate the power required for efficient conversion from the signal to the idler. We approximate the behavior of this partially constructive interference, assuming that the pump powers cancel the Kerr-induced phase modulation, by calculating the quality factor of the idler mode in FDTD simulation (see~\SMsec{sm:quality} for further details). This device, with parameters shared in Table~\ref {tab:optparams}, results in the intrinsic idler quality factor $Q^{cav}_{idl}=7.8$. 
\begin{table*}
    \renewcommand{\arraystretch}{1.4}   
    \setlength{\tabcolsep}{6pt}  
    \centering
    \begin{tabular}{|c|c|c|c|c|} 
    \hline
        Ring Parameters &$r$ & $w$ & $t_{cav}$ & $d_{ref}$\\
     \hline
        ($\mu m$) &$6$ & $0.75$ & $0.55$ & $0.87$\\
    \hline
    \hline
        Resonant Modes & Signal & Pump A & Pump B & Idler\\
     \hline
         m &$143$ & $28$ & $115$ & $0$\\
     \hline
         $\lambda$ ($\mu m$) &$0.615$ & $2.095$ & $0.750$ & $1.301$\\
    \hline
    \end{tabular}
    \\[0.2em]
    \begin{tabular}{|c|c|c|} 
    \hline
        Idler Results & $Q^{cav}_{idl}$ & $V_{cav}$\\
     \hline
        &$7.8$ & $7.3\cdot10^{-19}$  $m^3$\\
    \hline
    \end{tabular}
    \caption{Parameters and modes of the ring resonator}
    \label{tab:optparams}
\end{table*}
%

\section{Hamiltonian for Idler Efficiency}\label{m_Hamil}
To determine $\eta_{idler}$, we refer to the non-Hermitian effective Hamiltonian for a FWM process,
\begin{equation}
\begin{aligned} 
H=H_{channel}+H_{cavity}+H_{coupling}+H_{loss},
\end{aligned}
\end{equation}
where $H_{channel}$ describes the propagating modes that couple in to the cavity, $H_{cavity}$ represents the interactions and modes found in the cavity, $H_{coupling}$ determines the coupling between cavity and spatial channels, and $H_{loss}$ determines how light dissipates from the emitter and the cavity modes~\cite{Vernon2015, Vernon2016}. 

Relevant to the Hamiltonian are the internal, external, and loaded quality factors of each mode $Q_J^{int}$, $Q_J^{ext}$, and $Q_J=(1/Q_J^{int}+1/Q_J^{ext})^{-1}$, as well as the intrinsic quality factor of the cavity, $Q_J^{cav}=\omega_J/(2\Gamma_J^{cav})$, which is related to the intrinsic cavity loss rate $\Gamma_J^{cav}$. We also consider the bus coupling rate $\Gamma_J^{bus}=|\gamma|^2/(2v_J)=\alpha_J\Gamma_J^{cav}$ with waveguide coupling strength $\gamma_J$ and group velocity $v_J$~\cite{Vernon2016}. The parameter $\alpha_J$ allows us to model the waveguide coupling rate as a ratio of the intrinsic cavity loss, with critical coupling at $\alpha_J=1$. For our evaluation, we select $\alpha_{A}=1$, $\alpha_{B}=1$, and $\alpha_{sig}=0$.

These give rise to the associated linear channel-cavity coupling rate $\Gamma_J=\omega_J/(2Q_J^{ext})$ and the cavity loss rate $M_J=\omega_J/(2Q_J^{int})$ for each frequency. We can use these to calculate the total damping rate $\overline{\Gamma}_J=\Gamma_J+M_J$. For the pump modes, $\Gamma_J=\Gamma_J^{bus}=\alpha_J\Gamma_J^{cav}$, and $M_J=\Gamma_J^{cav}$, where $Q_J^{cav}$ is the intrinsic radiative loss of the cavity mode. Because we are designing around nonlinear coupling, there is no desirable linear output channel from the cavity at the signal frequency, thus $\Gamma_{sig}=0$ and $M_{sig}=\Gamma_{sig}^{bus}+\Gamma_{sig}^{cav}=(1+\alpha_{sig})\Gamma_{sig}^{cav}$.

In contrast, intrinsic radiation from the idler mode represents the successfully retrieved idler wave in the far-field. This rate is $\Gamma_{idl}=\Gamma_{idl}^{cav}$. The idler mode $m_{idl}=0$ is so weakly confined that $\Gamma_{idl}$ is much greater than any other loss, so $M_{idl}\simeq0$. This effectively transfers losses away from the idler mode by fully utilizing the radiation that is normally considered a loss. All of the cavity coupling rates mentioned above are summarized in~\ref{tab:rates}.
\begin{table*}
    \renewcommand{\arraystretch}{1.4}   
    \setlength{\tabcolsep}{6pt}  
    \centering
    \begin{tabular}{|c|c|c|c|c|} 
    \hline
        Cavity Mode (J) & Signal (sig) & Pump A (A) & Pump B (B) & Idler (idl)\\
     \hline
        $\Gamma_J$ (Units of $\Gamma_J^{cav}$) & 0 & $\alpha_A$ & $\alpha_B$ & $1$\\
     \hline
        $M_J$ (Units of $\Gamma_J^{cav}$) &$(1+\alpha_{sig})$ & $1$ & $1$ & $0$\\
    \hline
    \end{tabular}
    \caption{Relevant linear coupling rates to the system Hamiltonian. Note that when $\alpha_A=\alpha_B=1$, the coupling rate equals the loss rate for the pumps, leading to critical coupling.}
    \label{tab:rates}
\end{table*}

Entanglement protocols require that the color center emits a single photon. Conversion of a single photon destroys one photon from pump A and adds one photon to pump B. Because this transaction is negligible compared to the high pump power ($>1$ W), pump depletion effects will not affect the evolution of the Hamiltonian. Thus, we consider the pumps to be classical waves and drop their evolutionary terms from the channel Hamiltonian, keeping only their effect on the coupling between signal and idler. $\Gamma_{sig}=0$, and $\Gamma_{idl}$ belongs to the loss portion of the Hamiltonian, so
\begin{equation}
\begin{aligned} 
H_{channel}=0.
\end{aligned}
\end{equation}
We drop terms for the cavity modes $\sum_{J}\hbar\omega_{J}b^{\dagger}_{J}b_{J}$, where $b_J, b^\dagger_J$ are the annihilation and creation operators on the cavity resonances under the rotating wave approximation. Thus,
\begin{equation}
\begin{aligned} 
H_{cavity}=H_{NL},
\end{aligned}
\end{equation} where $H_{NL}$ is the nonlinear portion of the Hamiltonian that controls the transfer between signal and idler.
By assuming $\gamma\in\Re$, the classical amplitude of the pumps $a_A, a_B$ in the cavity given the input pump powers $P_A, P_B$ follows the relation
\begin{equation}
\begin{aligned} 
a_J=-\frac{ie^{i\xi}}{\omega_J}\sqrt{\frac{4\alpha_J P_JQ_J^{cav}}{\hbar(\alpha_J+1)^2}},
\end{aligned}
\end{equation}
where $\xi$ is the constant phase associated with the pump. The nonlinear Hamiltonian is given by the nonlinear coupling rate $g_{nl}=\Delta a_Aa_B^*$ and
\begin{equation}
\begin{aligned} 
H_{NL}=-\hbar g_{nl}[b_{sig}b^{\dagger}_{idl} + H. C.],
\end{aligned}
\end{equation}
\begin{equation}
\begin{aligned} 
\Delta=2\hbar\overline{\omega}^2cn_2/(\overline{n}^2V_{cav}),
\end{aligned}
\end{equation}
\begin{equation}
\begin{aligned} 
\overline{\omega}^2=\sqrt{\omega_{sig}\omega_{A}\omega_{B}\omega_{idl}},
\end{aligned}
\end{equation}
\begin{equation}
\begin{aligned} 
\overline{n}^2=\sqrt{n(\omega_{sig})n(\omega_{A})n(\omega_{B})n(\omega_{idl})}.
\end{aligned}
\end{equation}
where $b^\dagger_{J}$ and $b_{J}$ are the creation and lowering operators for the cavity modes, $\Delta$ is the nonlinear coupling strength parameter, $V_{cav}$ is the mode volume of the cavity, $n(\omega)$ is the linear refractive index of the material, and $n_2$ is the nonlinear refractive index of the material~\cite{Vernon2016}. 
The coupling portion of the Hamiltonian only contains the emitter to cavity coupling term, given the emitter's raising and lowering operators $\sigma_+$ and $\sigma_-$:
\begin{equation}
\begin{aligned} 
H_{coupling}=\hbar g_{e}(b_{sig}^{\dagger}\sigma_-+b_{sig}\sigma_+),
\end{aligned}
\end{equation}
\begin{equation}
\begin{aligned} 
g_e=\frac{1}{2}\sqrt{F_pM_{sig}\frac{2\pi r_{ZPL}}{\tau_e}},
\end{aligned}
\end{equation}
\begin{equation}
\begin{aligned} 
F_p=\frac{3}{4\pi}\left(\frac{\lambda_{sig}}{n(\omega_{sig})}\right)^3\frac{Q^{cav}_{sig}}{V_{cav}},
\end{aligned}
\end{equation}
where $g_e$ is the emitter to cavity coupling rate, $F_p$ is the Purcell enhancement of the signal mode, $\tau_e$ is the bulk radiative emission lifetime of the emitter. The ratio of ZPL radiative decay $r_{ZPL}=r_{Debye}\cdot r_{QE}$ is given by $r_{Debye}$, the Debye-Waller factor of the emitter, and $r_{QE}$, the quantum efficiency of the emitter due to non-radiative decay. This ratio is a vital term in determining the absolute efficiency of obtaining an entangled photon from the spin-photon interface~\cite{Brooks2021,Janitz:20}. The quantum efficiency and Debye-Waller factor for the SnV center are around 0.8 and 0.6 in bulk diamond, respectively, leading to a bulk radiative factor of $r_{ZPL}=0.48$~\cite{Ruf:2021, Iwasaki2017, Rosenthal2024}. This is relatively high compared to the SiV and NV centers, with radiative factors of around $r_{ZPL}=0.24, 0.04$, respectively, in bulk diamond~\cite{Redhu2025, Neu2011, Becker2017, Neu2012, Berthel2015}. We choose to design around the SnV center due to its high radiative factor, using $r_{ZPL}=0.48$ for our analysis.
The emitter loss rate is then $M_{e}=2\pi (1-r_{ZPL})/\tau_e$. 

\begin{figure*}
    \centering
    \includegraphics[width=0.8\linewidth]{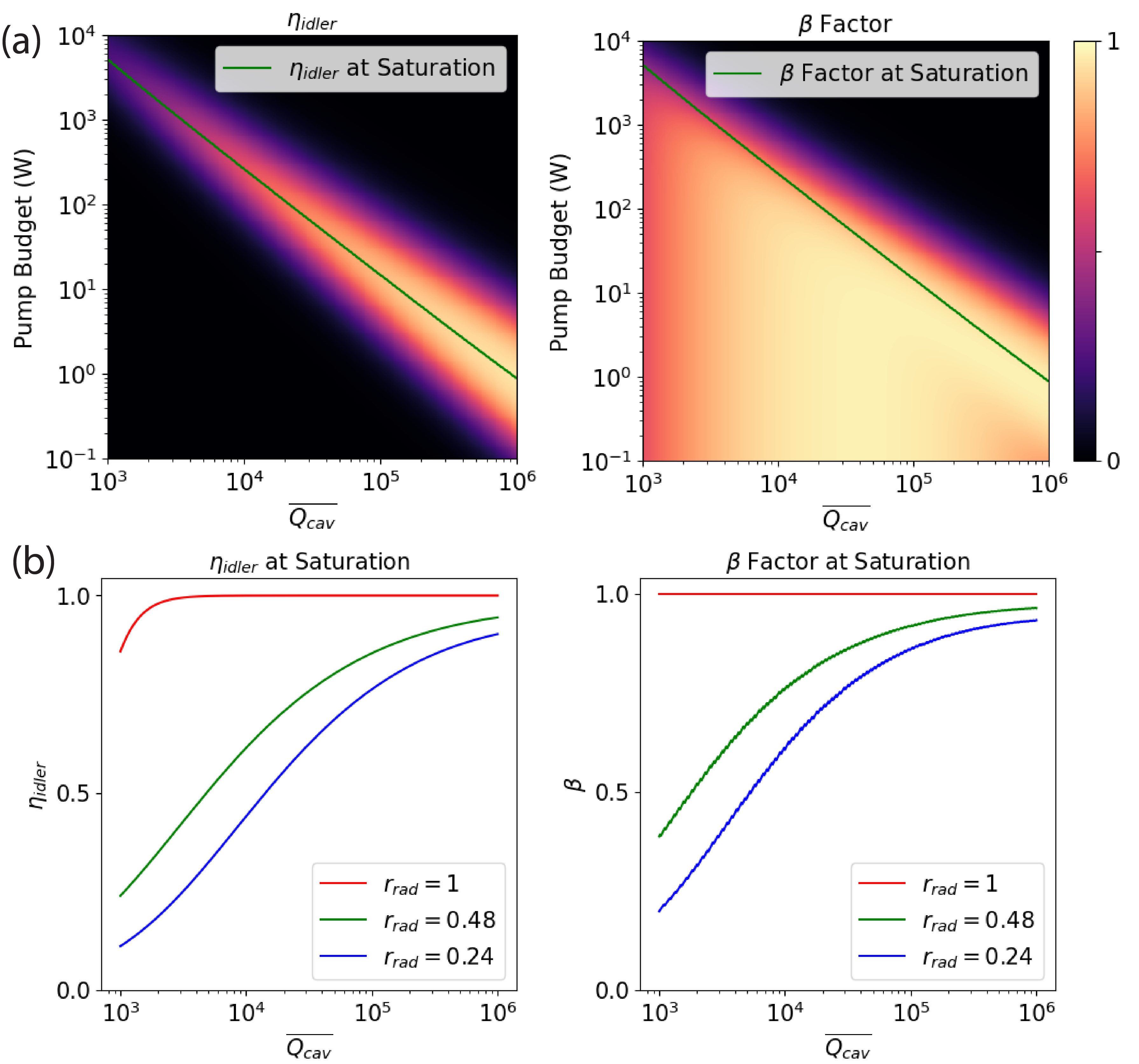}
    \caption{(a) $\eta_{idler}$ and $\beta$ factor plotted as a function of pump budget and $\overline{Q_{cav}}$ for $r_{ZPL}=0.48$. (b) The maximum $\eta_{idler}$ possible at saturation and corresponding $\beta$ factor, plotted for multiple values of $r_{ZPL}$, with a maximum pump budget of 10 kW. Note that $\eta_{idler}$ must be lower than the $\beta$ factor, demonstrating the need for high quality factor devices. }
    \label{fig:Fig4}
\end{figure*}

Putting together all sources of out-coupling from the system gives the non-Hermitian portion of the Hamiltonian,
\begin{equation}
\begin{aligned} 
H_{loss}=-i\hbar\left(\frac{\overline{\Gamma}_{sig}}{2}b_{sig}^{\dagger}b_{sig}+\frac{\Gamma_{idl}}{2}b_{idl}^{\dagger}b_{idl}+\frac{M_e}{2}\sigma_+\sigma_-\right)
\end{aligned}
\end{equation}
where we note that the accumulated radiative loss from the idler mode divided by the total accumulated loss of the system equals $\eta_{idler}$.
The Hamiltonian can be represented by a coupled amplitude equation with amplitudes $\mathbf{c}(t)=\begin{pmatrix}              
    c_{e}&c_{sig}&c_{idl}
\end{pmatrix}^T$ for the emitter, signal, and idler populations, respectively. The system
\begin{equation}
\begin{aligned} 
\mathbf{c}'(t)=-A\mathbf{c(t)},
\end{aligned}
\end{equation}
\begin{equation} 
\begin{aligned}
A= \begin{bmatrix} M_e/2 & ig_{e} & 0\\ ig_{e} & M_{sig}/2+i\Delta_{sig} & ig_{nl} \\ 0 & ig_{nl} & \Gamma_{idl}/2 + i\Delta_{idl} \end{bmatrix} 
\end{aligned} 
\end{equation} 
is a linear constant-coefficient homogeneous system of ordinary differential equations defined by the coefficient matrix $A$, where $\Delta_{sig}$ and $\Delta_{idl}$ are the detunings of the signal and idler frequencies from the resonance. This system has solutions given by the eigenvectors $V=\begin{pmatrix} \mathbf{v}^{(1)}&\mathbf{v}^{(2)}&\mathbf{v}^{(3)} \end{pmatrix}$ and corresponding eigenvalues $\lambda_j$. The exact solution in time is given by 
\begin{equation} 
\begin{aligned} \mathbf{c}(t)=\sum_{j=1}^{3}a_{j}e^{-\lambda_jt}v^{(j)}, 
\end{aligned} 
\end{equation} where $\mathbf{a}=V^{-1}\mathbf{c}(0)$. We numerically solve for 
\begin{equation} 
\begin{aligned} \eta_{idler}=\int_0^{\infty}\Gamma_{idl}|c_{idl}(t)|^2dt=\Gamma_{idl}\sum_{j,k=1}^3\frac{a_ja_k^*v_{3}^{(j)}(v_{3}^{(k)})^*}{\lambda_j+\lambda_k^*}, 
\end{aligned} 
\end{equation} 
\begin{equation} 
\begin{aligned} \beta=1-\eta_{emitter}=1-M_{e}\sum_{j,k=1}^3\frac{a_ja_k^*v_{1}^{(j)}(v_{1}^{(k)})^*}{\lambda_j+\lambda_k^*}, 
\end{aligned} 
\end{equation} 
where the $\beta$ factor represents the probability that a single decay leads to radiative ZPL emission, in contrast to lossy decay from the emitter $\eta_{emitter}$. In bulk diamond, $\beta=r_{ZPL}$, but in a cavity, we find that both the quality factor and $g_{nl}$ (influenced by pump power) lead to a change in $\beta$.
We plot the resulting $\eta_{idler}$ and $\beta$ factor vs pump budget $P_{budget}=2\sqrt{P_AP_B}$, where the factor of 2 accounts for injection of the pump from both bus waveguide ports, for selected intrinsic quality factors $\overline{Q^{cav}}=Q_{sig}^{cav}=Q_A^{cav}=Q_B^{cav}$ in Figure~\ref{fig:Fig4}. See \SMsec{sm:hamiltonian} for a comparison of eigenmode solver to Qutip simulation results, demonstrating a Root Mean Square Error (RMSE) of $2.5\cdot10^{-4}$. 

We note that $\eta_{idler}$ eventually decreases as more pump power is applied, shown in~\ref{fig:Fig4}a. This can be interpreted as the reduction in the $\beta$ factor of the emitter's coupling into the cavity modes, which occurs when the effective quality factor of the signal mode decreases due to high out-coupling from $g_{nl}$. The decrease in $\beta$ shows that lossy decay begins to dominate the emitter's output, causing these devices to have a maximum efficiency that is limited by the maximum absorption of light from the emitter into the cavity. We refer to the maximum achievable $\eta_{idler}$ for a chosen $\overline{Q^{cav}}$ as the saturation point, and plot it in~\ref{fig:Fig4}b for multiple values of $r_{ZPL}$. This reveals the relationship that smaller values of $r_{ZPL}$ lead not just to higher required pump budget, but also to a lower maximum efficiency.
%

\section{Discussion}\label{m_Discussion}
Using a pump budget of 15.2 W of on chip coupled power and assuming an overall cavity quality factor of $\overline{Q^{cav}}=10^5$ leads to a saturated efficiency of $\eta_{idler}=85.3\%$. Realistically, coupling onto the chip will also incur losses, which should increase the power required by the laser by less than an order of magnitude. Pulsed quasi-CW lasers can provide powers in the kW level sufficient to power this system, and varying the duty cycle of the pulse can combat thermal effects with a low average power~\cite{mehrabad2025multi}. Thus, with reasonable powers for pulsed quasi-CW lasers and attainable quality factor, the total efficiency of the system can be up to $\eta=0.18$ directly from the color center to an optical fiber at 1301 nm. In contrast, with fully radiative coupling ($r_{ZPL}=1$), the same cavity model with $\overline{Q^{cav}}=10^5$ yields efficiencies of $\eta_{idler}=95.3\%$ and $\eta=0.21$ with a pump budget of 18.3 W.

What is notable about this system is not that the modes are lossless but rather that the losses are transferred away from the conversion of the idler's mode by the careful selection of $\Gamma_{idl}$ such that $M_{idl}\simeq 0$. This design effectively utilizes the loss of the idler mode as engineered coupling into the far-field, rather than introducing a separate source of out-coupling. 

Using an SLM after the S-wave-plate could change the output beam, which is polarized in $\hat{x}$, to overlap nearly perfectly with a Gaussian beam~\cite{Forbes:16}. This could bring the spatial efficiency to the limit of the numerical aperture collection and bulk optics transmission,  $\eta_{spatial}=0.73\cdot0.9=0.66$, leading to overall efficiencies of $\eta=0.56$ for $r_{ZPL}=0.48$, or $\eta=0.63$ for $r_{ZPL}=1$. An SLM could also steer the beam spatially or modify the beam to a higher-order Gaussian mode for spatial or modal multiplexing.

Replacing the nonlinear out-coupling $g_{nl}$ with waveguide coupling or other free space coupling techniques still leads to a corresponding reduction in the $\beta$ factor, as long as $r_{ZPL}\neq 1$. The saturation of the efficiency is similar to previous results obtained for FWM when the signal source is a waveguide instead of an emitter~\cite{Vernon2016}. Using the correct out-coupling is important to optimizing the efficiency of collection from an emitter~\cite{mccutcheon2009broadband}. Nonlinear coupling provides a distinct advantage of tunability compared to physically structured mode coupling: the optimal operating point for $\eta_{idler}$ can be found by sweeping the power of the pumps to obtain the best combination of $\beta$ and $g_{nl}$. We provide additional insight on the importance of the $\beta$ factor in \SMsec{sm:Weighing}.

Coupling to a lossy idler mode subsequently demands higher pump powers to build up the idler field. While the pump powers for high efficiency is limited by the quality factor of the resonator modes, fabrication methods in diamond have the capability to achieve a $Q=10^5$, and quasi-static pulsed lasers have been shown to be effective in delivering high instantaneous power while maintaining low average power~\cite{Ding2024,flower2024observation,xu2025chip,mehrabad2025multi,mehrabad2025quantum}, leading us to believe that implementing this method is possible with current technologies. A diamond ring integrated onto another platform with a higher nonlinearity might reduce the power requirements, and could even utilize methods such as bound-state-in-continuum modes~\cite{ye2025sum}.

We also note that our method has the highest saturation efficiency when there is minimal coupling between the signal mode in the cavity and the bus waveguide ($\alpha_{sig}=0$). This can be achieved using a two-point coupling scheme that destructively interferes for the signal mode and couples well for the pump modes~\cite{Bo2017, Shoman:20}.

Given $Q^{cav}_{idl}=7.8$ at 1310 nm, we estimate the FPM condition is fulfilled for the idler cavity mode across 165 nm. However, the output wavelengths that can be generated by a single ring are limited to discretized solutions to the phase matching condition, as shown in Figure~\ref {fig:Fig2}b. By allowing for higher-order azimuthal phase in the idler mode number $m_{idl}$, which would correspond to higher-order Laguerre-Gaussian modes with radial polarization, the density of output wavelengths could be increased. An SLM could correct this azimuthal phase before passing through an S-wave plate. Further analysis into these modes and their quality factor is necessary in order to enable such a feature.

\section{Conclusion}\label{m_Conclusion}
We proposed a novel method for QFC that enables the collection of a $_{m_{idl}=0}$, low-Q mode from a direction normal to the propagation of the signal and pump waves, with output frequencies spanning 165 nm in the O-band. When applied to quantum networking, this method could be used to create contained unit cells for a qubit that interfaces directly with the infrared, reducing overall losses and increasing the scalability of color center qubits. Future work may be to experimentally confirm that added noise is negligible and single-photon purity under pulsed pumping and extend the scheme to other material stacks such as quantum dots in lithium niobate circuits~\cite{wang2025large}. Our spatio-spectral system architecture can enable compact, fabrication-ready building blocks for scalable telecom-band quantum networking architectures, beyond the limitations of SLMs or narrow-band and bulky fiber platforms~\cite{amorim2025spatial}.

We also presented an end-to-end analysis of the factors determining the efficiency of the spin-photon interface in a FFWM scheme, and found pump powers for which the $\beta$ factor and conversion to the idler mode lead to an optimal $\eta_{idler}$. By considering the non-radiative emission of the color center, we provide a more accurate model for the performance of color centers that demonstrates the impact of the quality factor and Purcell enhancement on spatio-spectral conversion. Continued system-level efficiency analysis of such integrated nonlinear photonic circuits~\cite{wu2025bidirectional} can pave the way forward to the improvement and realization of highly efficient spin-photon interfaces.
%

\section{Acknowledgements}
The authors wish to acknowledge fruitful discussions with Kartik Srinivasan and Ian Christian, as well as Sofia Patomaki, Yuqin Sophia Duan, Thomas Propson and Hamza Raniwala. This material is based upon work supported by the National Science Foundation Graduate Research Fellowship under Grant No. 2141064.


\section{Competing interests}

The authors declare no competing interests.

\section{Data and materials availability}

All of the data that support the findings of this study are reported in the main text and appendices. Source data are available from the corresponding authors on reasonable request.






  

  
  




\appendix
\renewcommand{\appendixname}{APPENDIX}

\newpage

\section{\MakeUppercase{fabrication considerations}}\label{sm:fab}
\begin{figure}[h]
    \centering
    \includegraphics[width=0.3\linewidth]{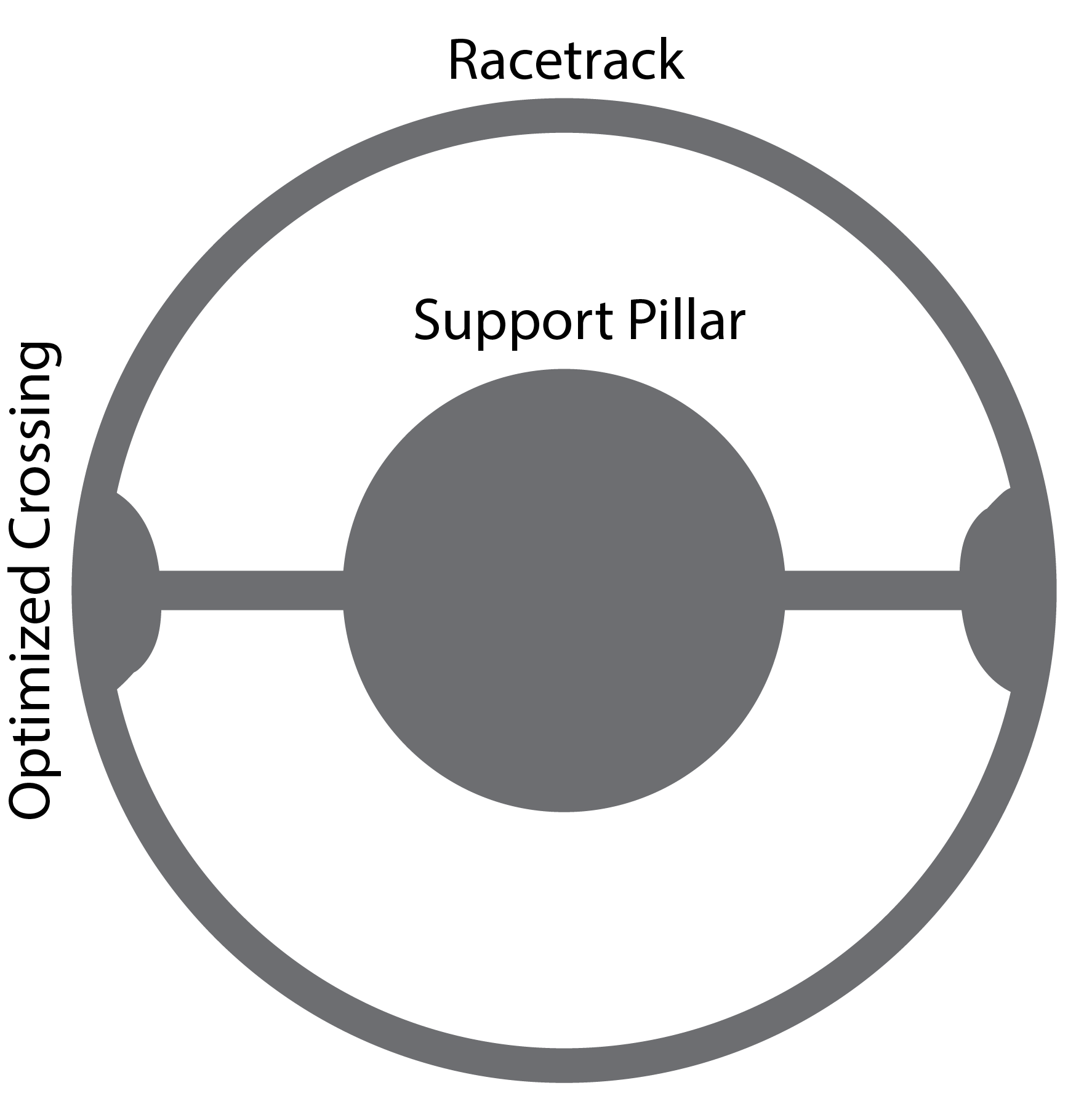}
    \caption{A potential fabrication scheme, displayed as a positive mask in diamond.}
    \label{fig:SupFig1}
\end{figure}

Microdisks have been fabricated with quality factors of up to $10^5$~\cite{Mitchell2019}. Ring resonators could be fabricated by providing inner supports that connect to a central pillar, as depicted in Figure~\ref{fig:SupFig1}. An undercut frees the outer ring, while remaining attached to the substrate via the central pillar. The crossing location between the supports and the ring can be optimized to reduce losses. Transferring this to a photonic chip with a waveguide could be performed with similar pick-and-place capabilities that already exist for diamond chiplets~\cite{Zhu2022, Starling2023, Pholsen:25}.
\section{\MakeUppercase{competing nonlinear noise processes}}\label{sm:noise}

We note two spontaneous four-wave mixing processes that could pollute the signal channel. In the first, two 750 nm photons are destroyed and create one 615 nm and one 961 nm photon. In the second noise process that fulfills energy conservation, one 750 nm and one 2095 nm photon is destroyed to create one 615 nm and one 5415 nm photon. However, we find through FDTD simulation that the 961 nm photon is detuned by ~1.8 (0.61) THz (FSRs) from the spatial mode that it couples to, and around ~1.2 (0.39) THz (FSRs) from the nearest resonance to which it is not phase matched. This significantly suppresses any field that might build up into the 961 nm mode, effectively negating the noise process. The nonlinear coupling strength also scales with frequency, meaning that the 5415 nm process will be significantly weaker than the main process. Using a weaker 750 nm pump and a stronger 2095 nm pump may also significantly reduce noise in the signal channel because the 750 nm pump is more likely to produce signal noise.

Noise processes are unlikely to pollute the idler channel. Any idler photon generated will have high momentum in the cavity plane, leading to greatly reduced overlap with the gaussian mode at the fiber, which must be reached by exiting the cavity plane.
 
As a final consideration, we note that the conditions given by 
\begin{equation}
\begin{aligned}
0=\omega_{sig}+\omega_A-\omega_B-\omega_{idl},
\end{aligned}
\end{equation}
\begin{equation}
\begin{aligned} 
0=m_{sig}+m_A-m_B
\end{aligned}
\end{equation}
also exists due to the symmetric injection of pumps; however, by phase matching according to the first set of FPM equations, this secondary condition is automatically phase-mismatched, and will be negligible compared to the main FFWM process.

\section{\MakeUppercase{ring mode profiles}}\label{sm:RingProfiles}
\begin{figure}[h]
    \centering
    \includegraphics[width=0.6\linewidth]{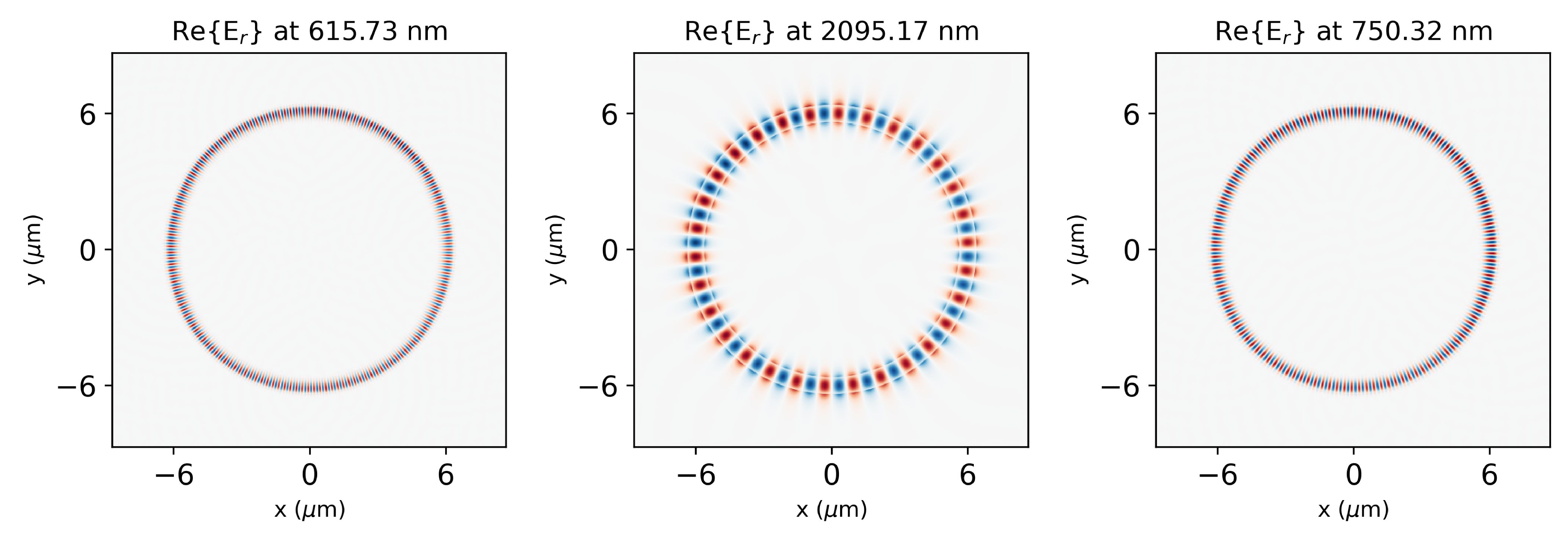}
    \caption{The real part of the radial component of the electric field, plotted for the signal, pump A, and pump B modes, respectively.}
    \label{fig:SupFig2}
\end{figure}

We obtain the electric field for the signal and pumps from FDTD simulation using Tidy3d~\cite{flores_tidy3d2026}. We use two passes to obtain the field: the first to extract resonances and quality factors, and the second to store a high density 3D grid of the electric field in the ring, shown in Figure~\ref{fig:SupFig2}. Because Maxwell's equations effectively model the propagation of a single photon in a spin-photon interface, we can calculate the idler mode from the $P^{(3)}$ these fields generate~\cite{duan2021vertically, li2016efficient, lu2019chip, lodahl2017chiral}.
\section{\MakeUppercase{low quality factor estimation}}\label{sm:quality}
Because the idler mode decays much faster than the duration of a source pulse, we cannot measure the quality factor using traditional FDTD methods that fit the decay of the resonance after the source finishes emitting. However, because $P^{(3)}$ completely decays by the end of a short simulation, we use the definition of quality factor given by~\cite{Haus_2004}
\begin{equation}
\begin{aligned} 
Q=\frac{\omega U}{P_d},
\end{aligned}
\end{equation}
where $U$ is the stored energy in the mode (measured by projecting a field monitor covering the mode volume onto the idler mode field), and $P_d$ is the power dissipated from the mode (measured as all flux out of the simulation).
\section{\MakeUppercase{spatial mode propagation}}\label{sm:Spatial}
\begin{figure}[h]
    \centering
    \includegraphics[width=0.5\linewidth]{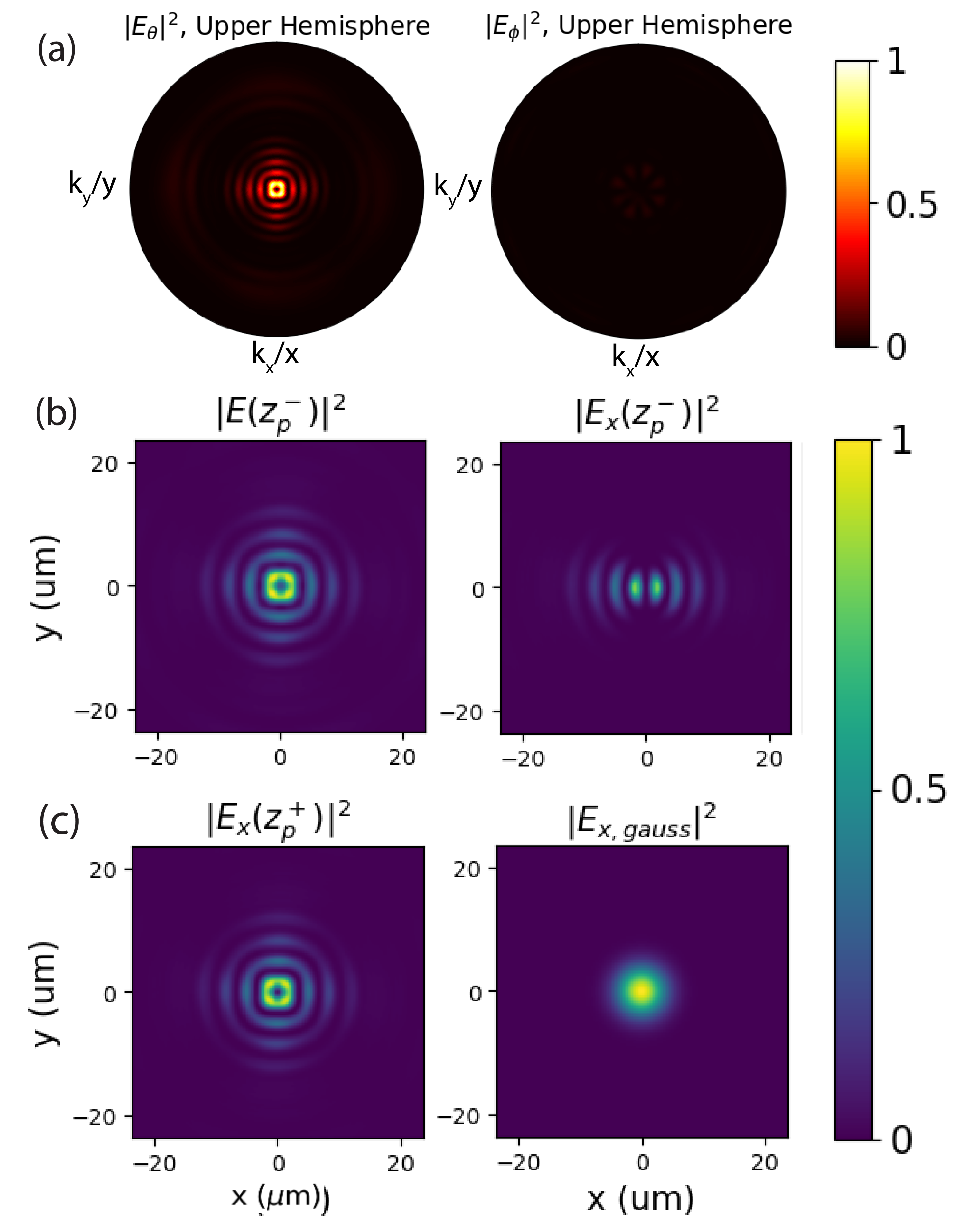}
    \caption{Spatial mode matching of the idler. (a) Far-field of the idler mode. (b) Transverse idler electric field profile calculated from Debye-Wolf integral, just before the S-waveplate. (c) Idler field $E_x(z_p^+)$ after S-waveplate transformation, next to its fitted Gaussian profile $E_{x, gauss}$.}
    \label{fig:SupFig3}
\end{figure}
Here we provide further depth on the stages of transformation used to compute the spatial overlap with a Gaussian beam, as depicted in Figure~\ref {fig:SupFig3}. After obtaining the far-field in Figure~\ref{fig:SupFig3}a, we apply the Debye-Wolf integral~\cite{Sherif:08}
\begin{equation} 
\begin{aligned}
\mathbf{E}(x_p, y_p, z_p)
= -\,\frac{ik_{0,idl}}{2\pi}
\iint_{\Theta}
\frac{\mathbf{a}(s_x, s_y)}{s_z}\,
\exp\!\Big[ ik \big( 
s_x x_p + s_y y_p + s_z z_p 
\big) \Big]\,
ds_x\, ds_y,
\end{aligned} 
\end{equation}
integrating over unit ray vector $\mathbf{s}$ the ray strengths $\mathbf{a}(s_x,s_y)$, resulting in the field in Figure~\ref{fig:SupFig3}b. We sweep values of $z_p$ near the focal plane, recording the electric field in the $xy$ plane, then apply the Jones matrix transformation for an S-waveplate~\cite{BhargavaRam:17},
\begin{equation} 
\begin{aligned}
\mathbf{E_{out}}(\rho, \phi, z_p)
=
\begin{bmatrix}
\cos\!\phi & \sin\!\phi \\
\sin\!\phi & -\cos\!\phi
\end{bmatrix}\mathbf{E}(\rho, \phi, z_p)^T
\end{aligned} 
\end{equation}
This results in an electric field that is nearly completely polarized in $E_x$. We plot this next to a Gaussian beam with waist size $w_0=4.6$~$\mu$m and a waist offset from the focal plane $\Delta z_p=-5$~$\mu$m in Figure~\ref{fig:SupFig3}c, resulting in the final power overlap of 24\%.

\section{\MakeUppercase{comparison of the eigenmode solver vs Qutip simulations}}\label{sm:hamiltonian}
\begin{figure}
    \centering
    \includegraphics[width=0.5\linewidth]{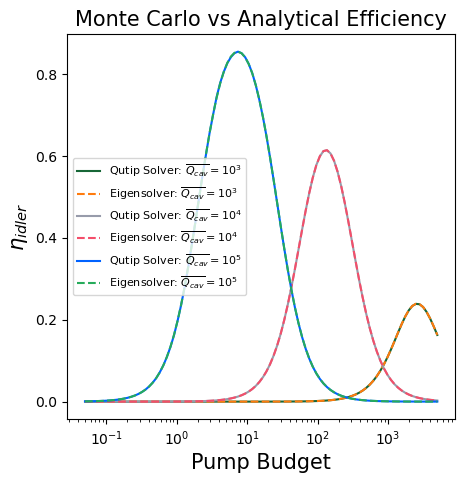}
    \caption{$\eta_{idler}$ calculated from the Qutip Solver and Eigenmode Solver for a range of pump budgets and $\overline{Q_{cav}}$}.
    \label{fig:SupFig4}
\end{figure}

We also solve the system Hamiltonian using Qutip Monte Carlo simulations, replacing effective loss terms with corresponding Linblad collapse operators, for a subset of Figure~\ref{fig:Fig4}a, and plot the comparison in Figure~\ref{fig:SupFig4}. We find nearly identical performance for $\eta_{idler}$ to the eigenmode solver with a Root Mean Square Error of $2.5\cdot10^{-4}$. This demonstrates that the effective Hamiltonian provides sufficient accuracy in modeling the full dynamics of this open quantum system.

\section{\MakeUppercase{weighing the importance of the emitter loss in the total loss}}\label{sm:Weighing}
\begin{figure}[h]
    \centering
    \includegraphics[width=0.4\linewidth]{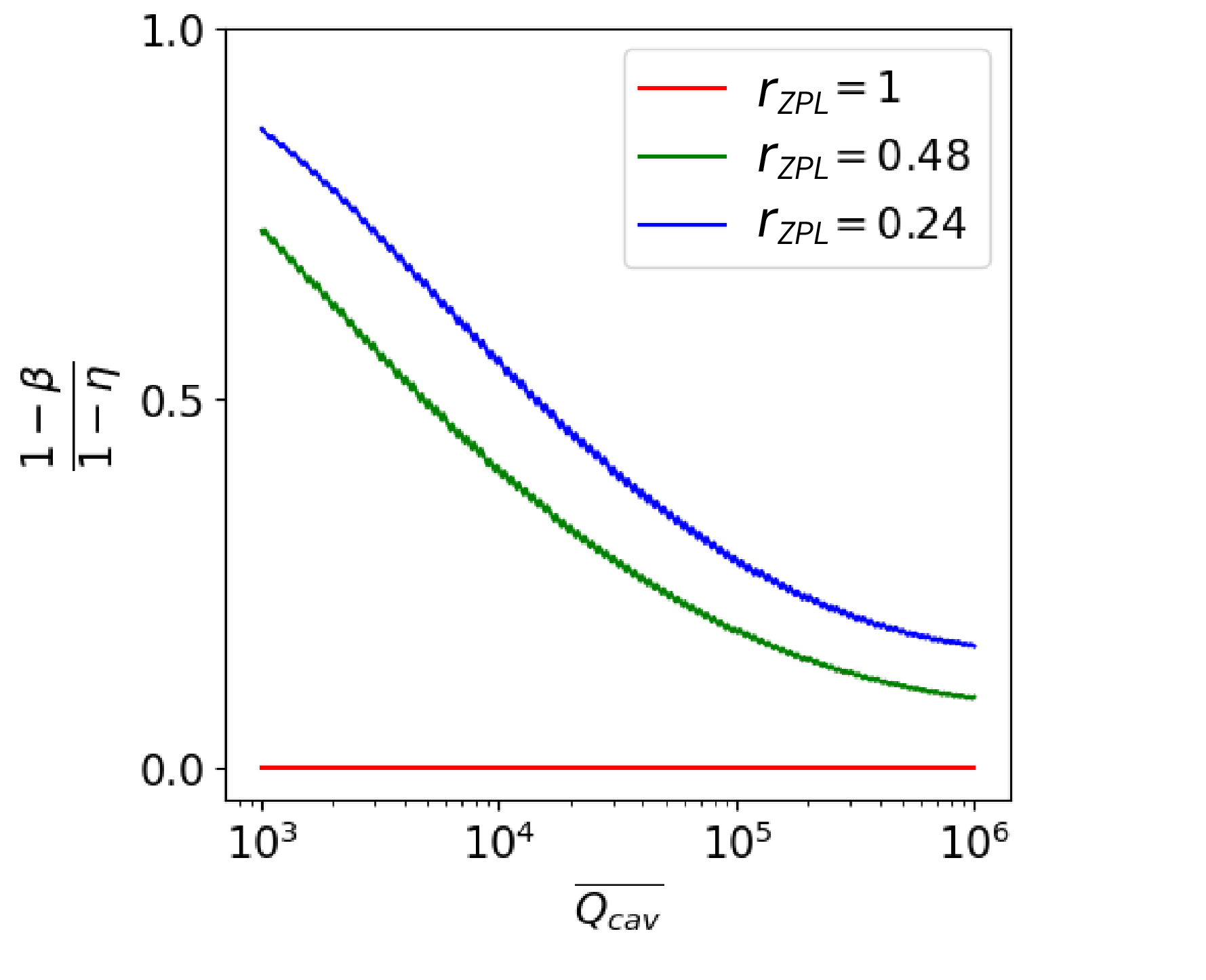}
    \caption{The ratio of emitter loss to total loss, plotted for several values of $r_{ZPL}$.}
    \label{fig:SupFig5}
\end{figure}

Here, we provide further depth to the effect of $r_{ZPL}$ on the efficiency limitations of emitters. We plot the ratio of emitter loss to total loss $\frac{1-\beta}{1-\eta}$ at saturation in Figure~\ref{fig:SupFig5}, assuming a SLM is used to achieve $\eta_{spatial}=0.66$. It is clear that the imperfect $\beta$ factor dominates the loss of this system for low quality factors, and only as the quality factor increases do the losses from $\eta_{spatial}$ begin to dominate. This illustrates the continued need for improvements to the Purcell enhancement in fabrication, as emitter losses become an even more dominant source of loss.
\newpage
\nocite{apsrev41Control}
\bibliography{Main}
\end{document}